\documentclass[preprint,times]{elsarticle}
\usepackage{amsmath}
\usepackage{mathrsfs}
\usepackage{amsthm}
\usepackage{epstopdf}
\usepackage{graphicx}
\journal{Optics Communications}
\begin{document}
\begin{frontmatter}
\title{Cross-spectral purity of nonstationary vector optical fields: A similarity with stationary fields}
\author{Rajneesh Joshi, \textsuperscript{1} and  Bhaskar Kanseri \textsuperscript{2*}}
\address{\textsuperscript{1} Department of Physics, Laxman Singh Mahar Campus, Pithoragarh-262502, Soban Singh Jeena University Almora, Uttarakhand, India \\\textsuperscript{2}Experimental Quantum Interferometry and Polarization (EQUIP), Department of Physics, Indian Institute of Technology Delhi, Hauz Khas, New Delhi-110016, India\\ *Corresponding author:- bkanseri@physics.iitd.ac.in}
\begin{abstract}
This study establishes a reduction formula for nonstationary cross-spectrally pure vector light fields with any spectral bandwidth. The formation of a reduction formula, analogous to that for stationary fields, does not apply to the normalized two-time Stokes parameters of a nonstationary field that is cross-spectrally pure. The current formula incorporates time-integrated coherence parameters to ensure cross-spectral purity. The reduction formula derived for nonstationary vector light fields with arbitrary spectral bandwidth shares a similar mathematical structure to that of reduction formulas used for stationary vector fields. Additionally, we examine the requirement of strict cross-spectral purity for using a time-integrated coherence function, which exhibits a mathematical expression similar to that of strict cross-spectral purity in stationary vector fields. This investigation sheds light on the cross-spectral purity of pulse-type fields, which holds potential applications in the field of statistical optics.       
\end{abstract}
\end{frontmatter}
\section{Introduction}
Cross-spectral purity (CSP) is a fundamental concept in classical optics that is extensively studied using Young's type interferometer (YI) in the case of stationary light fields \cite{mandel61, mandel95}. When the identical normalized spectra of light at two spatial input points, denoted as $\textbf{r}_1$ and $\textbf{r}_2$, interfere within the YI, the resulting normalized spectrum on the observation plane differs from the input spectra, indicating cross-spectrally impure fields \cite{mandel61}. However, under specific mathematical conditions, the normalized spectra at input points $\textbf{r}_1$ and $\textbf{r}_2$ can resemble the normalized spectra at a particular point $\textbf{R}$ on the observation plane, representing cross-spectrally pure optical fields \cite{mandel61}.

L. Mandel initially proposed CSP for stationary scalar light fields in 1961, where he observed an intriguing reduction in the complex degree of coherence (DOC) \cite{mandel61}. Subsequently, it was discovered that the absolute value of the complex spectral DOC remains the same across all frequencies of the optical field  \cite{mandel76}. Various techniques have been explored to investigate CSP in stationary scalar fields \cite{james97, kandpal93, wolf83, kandpal02,  friberg95, kanseri010, kanseri0010}, and the phenomenon has also been studied in the context of scattering and ghost imaging \cite{lahiri14, liu07}.

Subsequently, the concept of CSP was investigated for stationary vector light fields, specifically electromagnetic (EM) fields, considering all the polarization Stokes parameters \cite{hassinen09, friberg11}. Similar to the scalar case, a reduction formula exists for vector fields. In other words, the normalized space-time coherence Stokes parameters, divided by the corresponding usual Stokes parameters, can be expressed as the product of spatial and temporal coherence functions. Additionally, the absolute value of the normalized space-frequency coherence Stokes parameters, divided by the corresponding usual Stokes parameters, remains constant across all frequencies of the optical field.
Strict CSP is another important aspect of optical fields \cite{friberg11, peng17}. When all the Stokes parameters exhibit CSP at a single point on the observation plane of the YI, it is referred to as strict CSP. The equivalence between the space-time and space-frequency electromagnetic degree of coherence (EMDOC) and the degree of cross-polarization (DOCP) serves as a hallmark of strict CSP \cite{friberg11,joshi23}. Various approaches have also been explored in investigating CSP for stationary vector fields \cite{joshi23, hassinen13, chen14, partanen18}.

\section{CSP of nonstationary light fields}  
The most general form of the light field in optics is nonstationary or pulsed-like optical fields. Previous studies have examined their fundamental properties \cite{christov86, lazunen05, voipio13, koivurova019, ding20, lahiri10}. These fields exhibit characteristics related to coherence and polarization, which are influenced by specific time instants ($t_1$, $t_2$) or frequencies ($\omega_1$, $\omega_2$), as well as the time difference ($\Delta t$). A wavefront folding interferometer (WFI) \cite{guo18, bates46} and scanning wavefront folding interferometers \cite{koivura19} are used instead of YI in studying CSP for nonstationary light fields to avoid the frequency-dependent factors and spectral interference in YI.
The concept of CSP was initially introduced by M. Koivurove et al. in 2019 for nonstationary scalar light fields \cite{PRA19}. In CSP, the normalized spectral density remains the same at input spatial points $\textbf{r}_1$, $\textbf{r}_2$, and a point $\textbf{R}$ on the observation plane of the WFI. Furthermore, the two-frequency complex DOC should also be identical at $\textbf{r}_1$, $\textbf{r}_2$, and $\textbf{R}$. These conditions can be satisfied if the two-frequency complex DOC can be expressed as the product of space and frequency correlation functions. Similarly, the two-time complex DOC can be expressed as the product of space and time correlation functions.

Later, the phenomenon of CSP was expanded to nonstationary vector fields in both the domains of space-time and space-frequency \cite{joshi21}. The normalized Stokes parameters for CSP exhibit identical values at positions $\textbf{r}_1$, $\textbf{r}_2$, and the point $\textbf{R}$. Moreover, the two-frequency Stokes parameters normalized by their corresponding usual Stokes parameters also share the same values at $\textbf{r}_1$, $\textbf{r}_2$, and $\textbf{R}$. These criteria are met when the two-frequency Stokes parameters, normalized by their corresponding usual Stokes parameters, can be represented as a product of spatial and frequency correlation functions. Similarly, the two-time Stokes parameters, normalized by their corresponding usual Stokes parameters, are multiples of spatial and temporal correlation functions. The idea of strict CSP, originally developed for stationary vector fields, is expanded to nonstationary vector fields in the context of CSP \cite{joshi21}. If the two-frequency Stokes parameters, when normalized by the zeroth usual Stokes parameters, remain independent of frequency, and the two-time Stokes parameters, normalized by the zeroth usual Stokes parameters, remain independent of time, then it can be stated equivalently that the spatial components of both parameters are equal. This condition is referred to as the strict CSP of nonstationary vector fields \cite{joshi21}. Unlike stationary vector fields, strict CSP is only observed at a time difference of zero for nonstationary vector fields. \cite{joshi21}. The mathematical characteristics of the coherence function in CSP for nonstationary scalar and vector fields differ from those observed in stationary fields \cite{PRA19, joshi21}. However, a recent development involves deriving a reduction formula for the coherence function of nonstationary scalar fields by incorporating the time-integrated coherence function, which exhibits similarities to the results obtained for stationary scalar fields \cite{friberg23}. This advancement enables the extension of CSP to the EM domain, where polarization properties can be considered. 

In this paper, we examine the existence of a reduction formula for the CSP of nonstationary vector fields, which encompasses all the Stokes parameters. Our approach involves making use of time-integrated coherence functions. Furthermore, a condition for strict CSP involving time-integrated quantities is derived, exhibiting a mathematical form akin to the strict CSP observed in stationary vector fields. This is a novel approach to finding the conditions of CSP and strict CSP for nonstationary (modulated and fluctuating light beams as well as ultrashort laser pulses) vector fields, which are the most general fields and have numerous applications in science \cite{mandal20, wagner22}.
\begin{figure}[htbp]
\centering
\includegraphics[width=.8\linewidth]{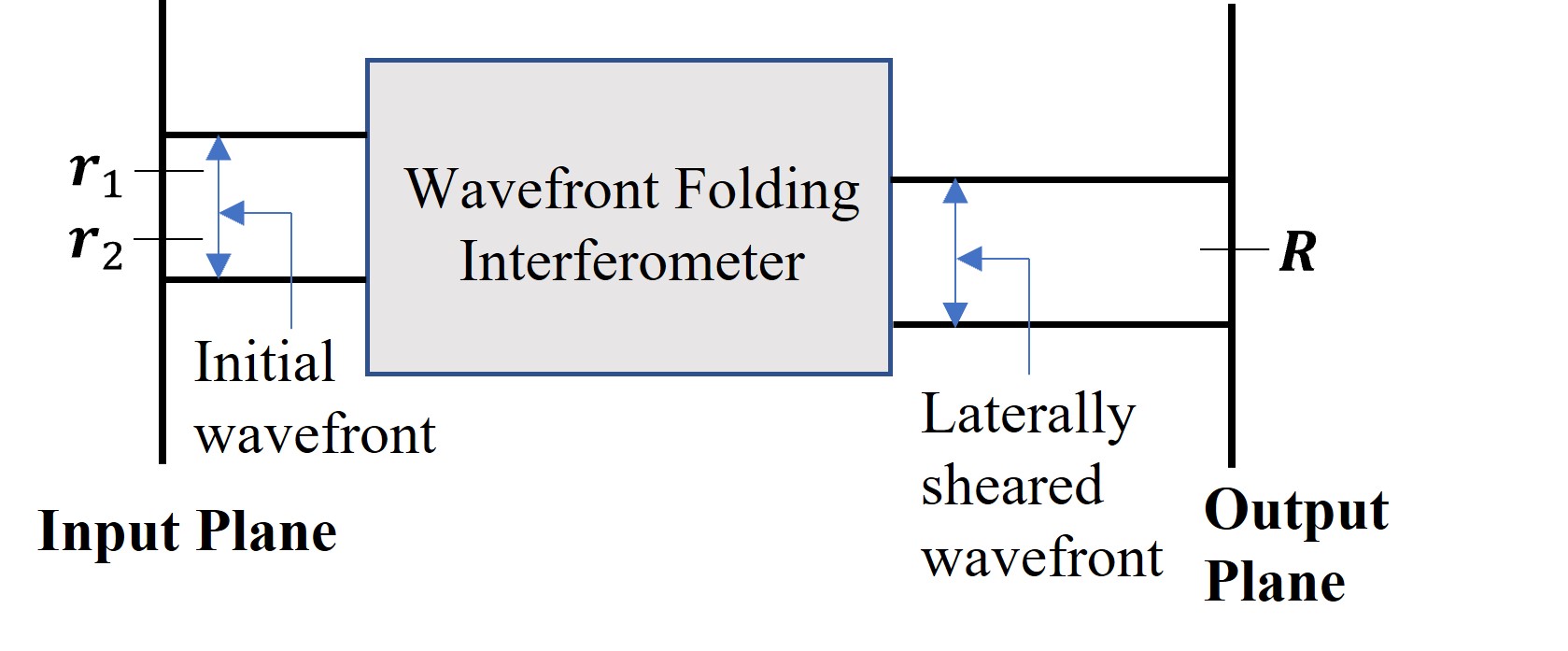}
\caption{Schematic configuration to study the cross-spectral purity of nonstationary vector field, $\textbf{r}_1$, $\textbf{r}_2$ are spatial input points, and $\textbf{R}$ is a point on the output plane. The interferometer produces a laterally sheared wavefront of the initial wavefront.}
\label{fig:false-color}
\end{figure} 
\section{CSP of Stokes parameters}
We consider the superposition of EM light fields through a WFI as shown in Fig. 1. If two EM fields at spatial input positions
$\textbf{r}_1$ and $\textbf{r}_2$ interfere, the net output field at point $\textbf{R}$ and time t can be expressed as \cite{joshi21} 
\begin{equation}
\begin{bmatrix}
E_x(\textbf{R},t) \\ E_y(\textbf{R},t)
\end{bmatrix}=\begin{bmatrix}
E_x(\textbf{r}_1,t-\tau_1) \\ E_y(\textbf{r}_1,t-\tau_1)\end{bmatrix}+\begin{bmatrix}
E_x(\textbf{r}_2,t-\tau_2) \\ E_y(\textbf{r}_2,t-\tau_2) 
\end{bmatrix},
\end{equation}
where $\tau_1=\frac{l_1}{c}$, $\tau_2=\frac{l_2}{c}$ are the times of light to reach point $\textbf{R}$ from $\textbf{r}_1$ and $\textbf{r}_2$, respectively. The distances $l_1$ and $l_2$ represent the measurements between point $\textbf{R}$ and the input points $\textbf{r}_1$ and $\textbf{r}_2$, respectively.
In the space-frequency domain, the net field at the output of WFI can be expressed as the combination of fields at two spatial positions in the input. The output field becomes \cite{joshi21}
\begin{equation}
\begin{bmatrix}
E_x(\textbf{R},\omega) \\ E_y(\textbf{R},\omega)
\end{bmatrix}=\begin{bmatrix}
E_x(\textbf{r}_1,\omega) \\ E_y(\textbf{r}_1,\omega)\end{bmatrix}\exp(\iota\omega\tau_1)+\begin{bmatrix}
E_x(\textbf{r}_2,\omega) \\ E_y(\textbf{r}_2,\omega) 
\end{bmatrix} \exp(\iota\omega\tau_2),
\end{equation}
where $\iota$ (iota) represents the imaginary number, $\omega$ denotes the angular frequency.
Assuming that the input spectral EM field is given as 
\begin{equation}
 E_i(\textbf{r},\omega)=E_{0_i}(\textbf{r},\omega) \exp[\iota\phi_i(\textbf{r},\omega)], (i=x,y),   
\end{equation}
where $E_{0_i}(\textbf{r},\omega)$ and $\phi_i(\textbf{r},\omega)$ denote the complex amplitude and phase of the optical field, respectively. 
The form of the wavefront phase is as $\phi_i(\textbf{r},\omega)=K\phi_i(r)$, $k=\frac{\omega}{c}$ is wave vector  and $\phi_i(r)$ represents the shape of the wavefront \cite{friberg23}. Finally, from this relation, we readily find that 
\begin{equation}
\phi_i(\textbf{r},\omega)=\frac{\omega}{\omega_0}\phi_i(\textbf{r},\omega_0), (i=x,y),    
\end{equation}
where $\omega_0$ denotes a fixed peak frequency of the spectrum.
Using Eqs. (3) and (4), Eq. (2)  takes the form
\begin{equation}
E_i(\textbf{R},\omega)=[E_{0_i}(\textbf{r}_1,\omega)+E_{0_i}(\textbf{r}_2,\omega)\exp(\iota\omega\tau_i(\textbf{r}_1,\textbf{r}_2))]\exp[\iota\omega(\tau_1+\frac{\phi_i(\textbf{r}_1,\omega_0)}{\omega_0})],   \end{equation}
where
\begin{subequations}
\begin{equation}
\tau_i(\textbf{r}_1,\textbf{r}_2)=\frac{\phi_i(\textbf{r}_2,\omega_0)-\phi_i(\textbf{r}_1,\omega_0)}{\omega_0}+\tau_2-\tau_1, \end{equation}
\begin{equation}
\tau_i(\textbf{r}_1,\textbf{r}_2)=(\tau_0)_i(\textbf{r}_1,\textbf{r}_2)+\Delta\tau.   
\end{equation}
\end{subequations}
As previously discussed, there are two conditions for the CSP of nonstationary vector fields. However, our findings and analysis focus solely on the first condition of CSP. Specifically, for nonstationary vector light fields, the first condition of CSP for the Stokes parameters can be described as follows:\cite{joshi21}
\begin{equation}
s_j(\textbf{r}_1,\omega)=s_j(\textbf{r}_2,\omega)=s_j(\textbf{R},\omega), (j=0-3),    
\end{equation}
where

\begin{equation}
s_j(\textbf{r},\omega)=\frac{S_{j}(\textbf{r},\omega)}{\int_{0}^{\infty}S_j(\textbf{r},\omega) d\omega} \end{equation}

are the Stokes parameters normalized with the corresponding usual Stokes parameters. Keeping in mind that the spectral Stokes parameters $S_j(\textbf{r},\omega)$ are defined as follows: \cite{mandel95} 
\begin{subequations}
\begin{equation}
S_0(\textbf{r},\omega)=\langle E^*_{x}(\textbf{r},\omega)E_{x}(\textbf{r},\omega)\rangle+\langle E^*_{y}(\textbf{r},\omega)E_{y}(\textbf{r},\omega)\rangle,
\end{equation}
\begin{equation}
S_1(\textbf{r},\omega)=\langle E^*_{x}(\textbf{r},\omega)E_{x}(\textbf{r},\omega)\rangle-\langle E^*_{y}(\textbf{r},\omega)E_{y}(\textbf{r},\omega)\rangle,
\end{equation}
\begin{equation}
S_2(\textbf{r},\omega)=\langle E^*_{y}(\textbf{r},\omega)E_{x}(\textbf{r},\omega)\rangle+\langle E^*_{x}(\textbf{r},\omega)E_{y}(\textbf{r},\omega)\rangle,
\end{equation}
\begin{equation}
S_3(\textbf{r},\omega)=\iota[\langle E^*_{y}(\textbf{r},\omega)E_{x}(\textbf{r},\omega)\rangle-\langle E^*_{x}(\textbf{r},\omega)E_{y}(\textbf{r},\omega)\rangle],
\end{equation}
\end{subequations}
here $\langle E^*_{i}(\textbf{r},\omega)E_{j}(\textbf{r},\omega)\rangle$, $(i,j=x,y)$ are the elements of the coherency matrix $J(\textbf{r},\omega)$. The asterisk and angular brackets in Eq. (9) indicate the complex conjugation and time average, respectively. 
By substituting the values of field components from Equation (5) into (9) and further assuming that both components of the EM field share the same phase, denoted as $\tau_x(\textbf{r}_1,\textbf{r}_2)=\tau_y(\textbf{r}_1,\textbf{r}_2)=\tau(\textbf{r}_1,\textbf{r}_2)$, we can determine the Stokes parameters at point $\textbf{R}$ as follows:
\begin{equation}
S_j(\textbf{R},\omega)=S_j(\textbf{r}_1,\omega)+S_j(\textbf{r}_2,\omega)+2Re[(S_j)_{0}(\textbf{r}_1,\textbf{r}_2,\omega,\omega)\exp(\iota\omega\tau(\textbf{r}_1,\textbf{r}_2))], (j=0-3),   
\end{equation}
where the two-frequency Stokes parameters in terms of polarization amplitudes are defined as \cite{voipio13}
\begin{subequations}
\begin{equation}
(S_{0})_{0}(\textbf{r}_1,\textbf{r}_2,\omega_1,\omega_2)=\langle E^*_{0_x}(\textbf{r}_1,\omega_1)E_{0_x}(\textbf{r}_2,\omega_2)\rangle+\langle E^*_{0_y}(\textbf{r}_1,\omega_1)E_{0_y}(\textbf{r}_2,\omega_2)\rangle,
\end{equation}
\begin{equation}
(S_1)_{0}(\textbf{r}_1,\textbf{r}_2,\omega_1,\omega_2)=\langle E^*_{0_x}(\textbf{r}_1,\omega_1)E_{0_x}(\textbf{r}_2,\omega_2)\rangle-\langle E^*_{0_y}(\textbf{r}_1,\omega_1)E_{0_y}(\textbf{r}_2,\omega_2)\rangle,
\end{equation}
\begin{equation}
(S_2)_0(\textbf{r}_1,\textbf{r}_2,\omega_1,\omega_2)=\langle E^*_{0_y}(\textbf{r}_1,\omega_1)E_{0_x}(\textbf{r}_2,\omega_2)\rangle+\langle E^*_{0_x}(\textbf{r}_1,\omega_1)E_{0_y}(\textbf{r}_2,\omega_2)\rangle,
\end{equation}
\begin{equation}
(S_3)_0(\textbf{r}_1,\textbf{r}_2,\omega_1,\omega_2)=\iota[\langle E^*_{0_y}(\textbf{r}_1,\omega_1)E_{0_x}(\textbf{r}_2,\omega_2)\rangle-\langle E^*_{0_x}(\textbf{r}_1,\omega_1)E_{0_y}(\textbf{r}_2,\omega_2)\rangle],
\end{equation}
\end{subequations}
where the elements of the two-frequency cross-spectral density matrix are represented by $(W_{ij})_0(\textbf{r}_1,\textbf{r}_2,\omega_1,\omega_2)=\langle E^*_{0_i}(\textbf{r}_1,\omega_1)E_{0_j}(\textbf{r}_2,\omega_2)\rangle$, $(i,j=x,y)$.
The first equality of the CSP condition in Eq. (7) can be expressed as follows: \cite{joshi21} 
\begin{equation}
S_j(\textbf{r}_2,\omega)=C(\textbf{r}_1,\textbf{r}_2)S_j(\textbf{r}_1,\omega),\end{equation}
where $C(\textbf{r}_1,\textbf{r}_2)$ is a frequency independent proportionality factor. Utilizing Eq. (12), Eq. (10) can be expressed as follows: 
\begin{equation}
S_j(\textbf{R},\omega)=S_j(\textbf{r}_1,\omega)[1+C(\textbf{r}_1,\textbf{r}_2)+2\sqrt{C(\textbf{r}_1,\textbf{r}_2)}Re[(\mu_j)_0(\textbf{r}_1,\textbf{r}_2,\omega,\omega)\exp(\iota\omega\tau(\textbf{r}_1,\textbf{r}_2))]],    
\end{equation}
where the definition of $(\mu_j)_0(\textbf{r}_1,\textbf{r}_2,\omega_1,\omega_2)$ can be expressed as \cite{joshi21}
\begin{equation}
(\mu_j)_0(\textbf{r}_1,\textbf{r}_2,\omega_1,\omega_2)=\frac{(S_j)_0(\textbf{r}_1,\textbf{r}_2,\omega_1,\omega_2)}{\sqrt{S_j(\textbf{r}_1,\omega_1)}\sqrt{S_j(\textbf{r}_2,\omega_2)}},   
\end{equation}
known as two-frequency Stokes parameters normalized by corresponding usual Stokes parameters, which is a complex quantity and can be represented as 
\begin{equation}
(\mu_j)_0(\textbf{r}_1,\textbf{r}_2,\omega_1,\omega_2)=|(\mu_j)_0(\textbf{r}_1,\textbf{r}_2,\omega_1,\omega_2)|\exp[\iota\alpha_j(\textbf{r}_1,\textbf{r}_2,\omega_1,\omega_2)],   \end{equation}
where $\alpha_j$ is phase of $(\mu_j)_0(\textbf{r}_1,\textbf{r}_2,\omega_1,\omega_2)$.
Applying Eq. (15) to (13), the Stokes parameters at point $\textbf{R}$ can be written as
\begin{equation}
\begin{split}
S_j(\textbf{R},\omega)&=S_j(\textbf{r}_1,\omega)[1+C(\textbf{r}_1,\textbf{r}_2)+2\sqrt{C(\textbf{r}_1,\textbf{r}_2)}\\&Re[|(\mu_j)_0(\textbf{r}_1,\textbf{r}_2,\omega,\omega)|\exp(\iota\alpha_j(\textbf{r}_1,\textbf{r}_2,\omega,\omega)+\iota\omega\tau(\textbf{r}_1,\textbf{r}_2))]]. 
\end{split}
\end{equation}
From the identity, $\exp(\iota\theta)=\cos\theta+\iota\sin\theta$, we readily find that
\begin{equation}
\begin{split}
S_j(\textbf{R},\omega)&=S_j(\textbf{r}_1,\omega)[1+C(\textbf{r}_1,\textbf{r}_2)+2\sqrt{C(\textbf{r}_1,\textbf{r}_2)}|(\mu_j)_0(\textbf{r}_1,\textbf{r}_2,\omega,\omega)|\\&\cos(\alpha_j(\textbf{r}_1,\textbf{r}_2,\omega,\omega)+\omega\tau(\textbf{r}_1,\textbf{r}_2))]. 
\end{split}
\end{equation}
By employing the second equality of the CSP condition, which states that $S_j(\textbf{r}_2,\omega)=S_j(\textbf{R},\omega)$, it becomes evident that the bracketed term in Eq. (13) must be frequency-independent. Consequently,
\begin{subequations}
\begin{equation}
(\mu_j)_0(\textbf{r}_1,\textbf{r}_2,\omega,\omega)\exp(\iota\omega\tau(\textbf{r}_1,\textbf{r}_2))=f(\textbf{r}_1,\textbf{r}_2,\Delta\tau),  \end{equation} 
\begin{equation}
(\mu_j)_0(\textbf{r}_1,\textbf{r}_2,\omega,\omega)\exp(\iota\omega[\tau_0(\textbf{r}_1,\textbf{r}_2)+\Delta\tau])=f(\textbf{r}_1,\textbf{r}_2,\Delta\tau).  
\end{equation}
\end{subequations}
Therefore, CSP exists at $\textbf{r}_1$ and $\textbf{r}_2$ if Eq. (12) holds together with Eq. (18). It is observed that $|(\mu_j)_0(\textbf{r}_1,\textbf{r}_2,\omega,\omega)|$ becomes frequency-independent. To achieve this, the choice of time delays $\Delta \tau$ should be such that the frequency-dependent term is eliminated in Eqs. (15) and (18) \cite{friberg23}. \newline
Examining coherence functions in both the space-time and space-frequency domains is an essential requirement in studying CSP. The two-time Stokes parameters are defined as follows: \cite{voipio13} 
\begin{subequations}
\begin{equation}
S_0(\textbf{r}_1,\textbf{r}_2,t_1,t_2)=\langle E^*_{x}(\textbf{r}_1,t_1)E_{x}(\textbf{r}_2,t_2)\rangle+\langle E^*_{y}(\textbf{r}_1,t_1)E_{y}(\textbf{r}_2,t_2)\rangle,
\end{equation}
\begin{equation}
S_1(\textbf{r}_1,\textbf{r}_2,t_1,t_2)=\langle E^*_{x}(\textbf{r}_1,t_1)E_{x}(\textbf{r}_2,t_2)\rangle-\langle E^*_{y}(\textbf{r}_1,t_1)E_{y}(\textbf{r}_2,t_2)\rangle,
\end{equation}
\begin{equation}
S_2(\textbf{r}_1,\textbf{r}_2,t_1,t_2)=\langle E^*_{y}(\textbf{r}_1,t_1)E_{x}(\textbf{r}_2,t_2)\rangle+\langle E^*_{x}(\textbf{r}_1,t_1)E_{y}(\textbf{r}_2,t_2)\rangle,
\end{equation}
\begin{equation}
S_3(\textbf{r}_1,\textbf{r}_2,t_1,t_2)=\iota[\langle E^*_{y}(\textbf{r}_1,t_1)E_{x}(\textbf{r}_2,t_2)\rangle-\langle E^*_{x}(\textbf{r}_1,t_1)E_{y}(\textbf{r}_2,t_2)\rangle],
\end{equation}
\end{subequations}
where the elements of the two-time mutual coherence matrix are represented by \newline$\Gamma_{ij}(\textbf{r}_1,\textbf{r}_2,t_1,t_2)=\langle E^*_{i}(\textbf{r}_1,t_1)E_{j}(\textbf{r}_2,t_2)\rangle$, $(i,j=x,y)$. 
Equation (24) in reference \cite{friberg23} can be extended for EM fields, and two-time Stokes parameters govern the complete information of coherence and polarization for EM fields \cite{voipio13}. The time-integrated version of two-point Stokes parameters for the EM field can be expressed as \cite{friberg23}
\begin{equation}
\Bar{S}_j(\textbf{r}_1,\textbf{r}_2,\Delta t)=\frac{1}{2\pi}\int_{-\infty}^{\infty}S_j(\textbf{r}_1,\textbf{r}_2,t,\Delta t)dt= \int_{0}^{\infty}S_j(\textbf{r}_1,\textbf{r}_2,\omega,\omega)\exp[-\iota\omega\Delta t]d\omega,   
\end{equation}
where $t=\frac{t_1+t_2}{2}$, and $\Delta t=t_2-t_1$.
Putting $\textbf{r}_1=\textbf{r}_2=\textbf{r}$, in Eq. (20) we obtain
\begin{equation}
\Bar{S}_j(\textbf{r},\textbf{r},\Delta t)=\frac{1}{2\pi}\int_{-\infty}^{\infty}S_j(\textbf{r},\textbf{r},t,\Delta t)dt= \int_{0}^{\infty}S_j(\textbf{r},\textbf{r},\omega,\omega)\exp[-\iota\omega\Delta t]d\omega.
\end{equation}
After straightforward developments of Eqs. (12), (14), and (20), we obtain 
\begin{equation}
\Bar{S}_j(\textbf{r}_1,\textbf{r}_2,\Delta t)=\sqrt{C(\textbf{r}_1,\textbf{r}_2)}\int_{0}^{\infty}S_j(\textbf{r}_1,\omega)\mu_j(\textbf{r}_1,\textbf{r}_2,\omega,\omega)\exp[-\iota\omega\Delta t]d\omega,   
\end{equation}
where the Eq. (14) provides the expression for $\mu_j(\textbf{r}_1,\textbf{r}_2,\omega,\omega)$ without any subscript. This quantity is associated with the electric field, as indicated by the left-hand side (LHS) of Eq. (3). 
On recalling Eqs. (3), (4), and (11) we obtain
\begin{equation}
S_j(\textbf{r}_1,\textbf{r}_2,\omega,\omega)=(S_j)_0(\textbf{r}_1,\textbf{r}_2,\omega,\omega)\exp[\iota\omega\tau_0(\textbf{r}_1,\textbf{r}_2)].
\end{equation}
Hence, the value of Eq. (14) yields
\begin{equation}
\mu_j(\textbf{r}_1,\textbf{r}_2,\omega,\omega)=(\mu_j)_0(\textbf{r}_1,\textbf{r}_2,\omega,\omega)\exp[\iota\omega\tau_0(\textbf{r}_1,\textbf{r}_2)].
\end{equation}
Inserting Eq. (18) into (24), the above equation can be written as 
\begin{equation}
\mu_j(\textbf{r}_1,\textbf{r}_2,\omega,\omega)=f(\textbf{r}_1,\textbf{r}_2,\Delta\tau)\exp[-\iota\omega\Delta\tau].
\end{equation}
By substituting the Eqs. (21) and (25) in Eq. (22), we obtain
\begin{equation}
\Bar{S}_j(\textbf{r}_1,\textbf{r}_2,\Delta t)=\sqrt{C(\textbf{r}_1,\textbf{r}_2)}f(\textbf{r}_1,\textbf{r}_2,\Delta \tau)\Bar{S}_j(\textbf{r}_1,\textbf{r}_1,\Delta t+\Delta \tau).   
\end{equation}
By employing Eqs. (12) and (21), we obtain
\begin{equation}
\Bar{S}_j(\textbf{r}_2,\textbf{r}_2,0)=C(\textbf{r}_1,\textbf{r}_2)\Bar{S}_j(\textbf{r}_1,\textbf{r}_1,0).    
\end{equation}
In analogy to the spectral Eq. (14), the definition of  two-time Stokes parameters normalized by corresponding usual Stokes parameters in terms of time-integrated correlations are expressed as \cite{joshi21}
\begin{equation}
\Bar{\psi}_j(\textbf{r}_1,\textbf{r}_2,\Delta t)=\frac{\Bar{S}_j(\textbf{r}_1,\textbf{r}_2,\Delta t)}{\sqrt{\Bar{S}_j(\textbf{r}_1,\textbf{r}_1,0)}\sqrt{\Bar{S}_j(\textbf{r}_2,\textbf{r}_2,0)}}.  \end{equation}
By putting the values from Eqs. (26) and (27), we get
\begin{equation}
\Bar{\psi}_j(\textbf{r}_1,\textbf{r}_2,\Delta t)=\frac{f(\textbf{r}_1,\textbf{r}_2,\Delta \tau)\Bar{S}_j(\textbf{r}_1,\textbf{r}_1,\Delta t+ \Delta \tau)}{\Bar{S}_j(\textbf{r}_1,\textbf{r}_1,0)}=f(\textbf{r}_1,\textbf{r}_2,\Delta \tau)\Bar{\psi}_j(\textbf{r}_1,\textbf{r}_1,\Delta t+\Delta \tau).    
\end{equation} 
By setting  $\Delta t=-\Delta \tau$ in Eq. (29) and observing that  $\Bar{\psi}_j(\textbf{r},\textbf{r},0)=1$ in Eq. (28), we can conclude that 
\begin{equation}
\Bar{\psi}_j(\textbf{r}_1,\textbf{r}_2,-\Delta \tau)=f(\textbf{r}_1,\textbf{r}_2,\Delta \tau).    
\end{equation}     
Therefore, Eq. (29) gives us the following result:
\begin{equation}
\Bar{\psi}_j(\textbf{r}_1,\textbf{r}_2,\Delta t)=\Bar{\psi}_j(\textbf{r}_1,\textbf{r}_2,-\Delta \tau)\Bar{\psi}_j(\textbf{r}_1,\textbf{r}_1, \Delta t+\Delta \tau).    
\end{equation}
This represents the reduction formula for nonstationary light fields, which is the key outcome of this paper. Equation (31) can be interpreted as the CSP of Stokes parameters for nonstationary vector fields. It reveals that the time-integrated spatiotemporal function can be expressed as a product of time-integrated spatial and temporal-dependent functions. This finding is analogous to the earlier discovery of CSP for stationary vector fields, expressed as $\psi_j(\textbf{r}_1,\textbf{r}_2,\tau)=\psi_j(\textbf{r}_1,\textbf{r}_2,\tau_j)\psi_j(\textbf{r}_1,\textbf{r}_1,\tau-\tau_j)$. \newline
Now, we can express the inverse form of Eq. (20) as follows:
\begin{equation}
S_j(\textbf{r}_1,\textbf{r}_2,\omega,\omega)=\frac{1}{2\pi}\int_{-\infty}^{\infty}\Bar{S}_j(\textbf{r}_1,\textbf{r}_2,\Delta t)\exp[\iota\omega\Delta t]d \Delta t.   
\end{equation}
By using Eqs. (4) and (21), we find that
\begin{equation}
\frac{S_j(\textbf{r},\omega)}{\Bar{S}_j(r,r,0)}=\frac{S_j(\textbf{r},\omega)}{\int_{0}^{\infty}S_j(\textbf{r},\omega)d\omega}=s_j(\textbf{r},\omega).    
\end{equation}
By inserting the  Eqs. (28) and (33) into (14), we obtain
\begin{equation}
\mu_j(\textbf{r}_1,\textbf{r}_2,\omega,\omega)=\frac{1}{2\pi} \frac{1}{\sqrt{s(\textbf{r}_1,\omega)s(\textbf{r}_2,\omega)}}\int_{-\infty}^{\infty}\Bar{\psi}_j(\textbf{r}_1,\textbf{r}_2,\Delta t)\exp(\iota\omega\Delta t)d\Delta t.   
\end{equation}
Using Eq. (28) and (33), we obtain as
\begin{equation}
\frac{\Bar{\psi}_j(\textbf{r},\textbf{r},\Delta t)}{s_j(\textbf{r},\omega)}=\frac{\Bar{S}_j(\textbf{r},\textbf{r},\Delta t)}{\Bar{S}_j(\textbf{r},\textbf{r},0) s_j(\textbf{r},\omega)}=\frac{\Bar{S}_j(\textbf{r},\textbf{r},\Delta t)}{S_j(\textbf{r},\omega)}.   
\end{equation}
Specifically, by utilizing Eqs. (31), (32), and (35), we can derive the expression for $\mu_j(\textbf{r}_1,\textbf{r}_2,\omega,\omega)$ as follows:
\begin{equation}
\mu_j(\textbf{r}_1,\textbf{r}_2,\omega,\omega)= \Bar{\psi}_j(\textbf{r}_1,\textbf{r}_2,-\Delta \tau) \exp(-\iota\omega\Delta \tau).  
\end{equation}
This demonstrates that the absolute value of spectral coherence Stokes parameters, normalized by their corresponding usual Stokes parameters, remain constant for all frequencies of the optical field, similar to the behavior observed in stationary vector fields. Therefore, by considering time-integrated quantities, it is possible to derive a reduction formula for nonstationary vector fields that exhibits similarities to the reduction formula observed in stationary vector fields.
\section{Strict CSP}
Now, let us direct our focus towards strict CSP in the context of nonstationary optical fields. Similar to the stationary cases, it is possible to derive an expression for strict CSP in nonstationary optical fields by incorporating time-integrated coherence functions. The two-frequency and two-time Stokes parameters, normalized by their respective zeroth Stokes parameters, are denoted as $\eta_j(\textbf{r}_1,\textbf{r}_2,\omega_1,\omega_2)$ and $\nu_j(\textbf{r}_1,\textbf{r}_2,t_1,t_2)$, respectively. These parameters are defined as follows:\cite{joshi21} 
\begin{subequations}
\begin{equation}
\eta_j(\textbf{r}_1,\textbf{r}_2,\omega_1,\omega_2)=\frac{S_j(\textbf{r}_1,\textbf{r}_2,\omega_1,\omega_2)}{\sqrt{S_0(\textbf{r}_1,\omega_1)S_0(\textbf{r}_2,\omega_2)}}, \end{equation}
\begin{equation}
\nu_j(\textbf{r}_1,\textbf{r}_2,t_1,t_2)=\frac{S_j(\textbf{r}_1,\textbf{r}_2,t_1,t_2)}{\sqrt{S_0(\textbf{r}_1,t_1)S_0(\textbf{r}_2,t_2)}}.
\end{equation}
\end{subequations} 
Developing Eq. (37) with Eq. (36) and utilizing Eq. (28) in reference \cite{joshi21}, we can write a relationship between the space-time and space-frequency coherence Stokes parameters normalized by zeroth Stokes parameters, which reads
\begin{equation}
\eta_j(\textbf{r}_1,\textbf{r}_2,\omega,\omega)= \Bar{\nu}_j(\textbf{r}_1,\textbf{r}_2,-\Delta \tau) \exp(-\iota\omega\Delta \tau).  
\end{equation} 
The form of this relationship bears a resemblance to $\eta_j(r_1,r_2,\omega)=\nu_j(r_1,r_2,\tau_0)\exp({\iota\omega\tau_0})$, which is observed in the case of stationary vector fields.
By introducing the EMDOC for nonstationary fields in space-frequency ($\mu_\epsilon$) and space-time ($\gamma_\epsilon$) domain via the formulas \cite{voipio13, setala006}
\begin{subequations}
\begin{equation}
\mu_\epsilon^2(\textbf{r}_1,\textbf{r}_2,\omega_1,\omega_2)=\frac{1}{2}\sum_{j=0}^{3}|\eta_j(\textbf{r}_1,\textbf{r}_2,\omega_1,\omega_2)|^2,   
\end{equation}
\begin{equation}
\gamma_\epsilon^2(\textbf{r}_1,\textbf{r}_2,t_1,t_2)=\frac{1}{2}\sum_{j=0}^{3}|\nu_j(\textbf{r}_1,\textbf{r}_2,t_1,t_2)|^2.   
\end{equation}
\end{subequations}
Putting the value from Eq. (38), yields the interesting result of
\begin{equation}
\mu_\epsilon^2(\textbf{r}_1,\textbf{r}_2,\omega,\omega)=\Bar{\gamma}_\epsilon^2(\textbf{r}_1,\textbf{r}_2,-\Delta \tau).    
\end{equation}
Therefore, the spectral EMDOC and the time-integrated version of space-time EMDOCs are equivalent.
Recently, a new condition has been introduced for strict CSP \cite{joshi23}, which involves the equality of the DOCP \cite{shirai07, hassinen11} in both the space-time and space-frequency domains. The EMDOC and DOCP are related to the equation in the following manner  \cite{joshi23}:
\begin{subequations}
\begin{equation}
\mu_{\epsilon}^2(\textbf{r}_1,\textbf{r}_2,\omega_1, \omega_2)=\frac{1}{2}|\mu_0(\textbf{r}_1,\textbf{r}_2,\omega_1,\omega_2)|^2[1+P^2(\textbf{r}_1,\textbf{r}_2,\omega_1, \omega_2)],    
\end{equation}
\begin{equation}
\gamma_{\epsilon}^2(\textbf{r}_1,\textbf{r}_2,t_1,t_2)=\frac{1}{2}|\psi_0(\textbf{r}_1,\textbf{r}_2,t_1,t_2)|^2[1+P^2(\textbf{r}_1,\textbf{r}_2,t_1,t_2)],    
\end{equation}
\end{subequations}
where the parameter $\mu_0(\textbf{r}_1,\textbf{r}_2,\omega_1,\omega_2)$ can be obtained from Eq. (14) (without the subscript) by setting $j=0$. Additionally, the parameter $\psi_0(\textbf{r}_1,\textbf{r}_2,t_1,t_2)=\frac{S_0(\textbf{r}_1,\textbf{r}_2,t_1,t_2)}{\sqrt{S_0(\textbf{r}_1,t_1)}\sqrt{S_0(\textbf{r}_2,t_2)}}$. $P(\textbf{r}_1,\textbf{r}_2,\omega_1, \omega_2)$ represents the two-frequency DOCP, while $P(\textbf{r}_1,\textbf{r}_2,t_1,t_2)$ corresponds to the two-time DOCP.    
By applying the Eqs. (36), (40), and (41), we obtain
\begin{equation}
P(\textbf{r}_1,\textbf{r}_2,\omega,\omega)=\Bar{P}(\textbf{r}_1,\textbf{r}_2,-\Delta\tau).
\end{equation}
Like EMDOC, the spectral DOCP and the time-integrated version of space-time DOCP are equivalent.
Equations (38), (40), and (42) represent the distinctive characteristics of strict CSP. These equations serve as the other key findings of this paper, highlighting the significance of strict CSP in nonstationary vector fields. 
\section{Conclusion}
In conclusion, this study establishes the conditions of cross-spectral purity and strict cross-spectral purity by incorporating time-integrated coherence parameters. The mathematical structure of time-integrated normalized two-time Stokes parameters exhibits similarity to the condition observed in stationary vector fields. The time-integrated version of normalized space-time coherence Stokes demonstrates a reduction property, which can be expressed as the product of spatial and temporal coherence functions. Furthermore, the absolute value of the normalized spectral coherence Stokes parameter remains consistent across all frequencies of the optical field. Similar to the strict CSP condition observed in stationary vector fields, we establish this condition for nonstationary vector fields as well. This is accomplished by ensuring that the EMDOCs and DOCPs in both the space-time and space-frequency domains are equal. Thus, our study highlights that the outcomes of cross-spectral purity in nonstationary vector fields align with those in stationary vector fields when considering time-integrated coherence functions. 
\section*{Disclosures}
 The authors declare no conflicts of interest.

\end{document}